\documentclass[3p,times,twocolumn]{elsarticle}

\usepackage{ecrc}

\newcommand*{\La}{\cal{L}}
\newcommand*{\no}{\noindent}
\newcommand*{\bea}{\begin{eqnarray}}
\newcommand*{\eea}{\end{eqnarray}}
\newcommand*{\be}{\begin{equation}}
\newcommand*{\ee}{\end{equation}}
\newcommand*{\pd}{\partial}
\newcommand*{\pdm}{\pd_{\mu}}
\newcommand*{\pdn}{\pd_{\nu}}

\newcommand*{\pref}[1]{(\ref{#1})}

\newcommand*{\mn}{{\mu\nu}}

\newcommand*{\nn}{\nonumber}

\newcommand*{\tr}{\mathrm{tr}}

\newcommand{\bma}{\begin{pmatrix}}
\newcommand{\ema}{\end{pmatrix}}

\volume{00}

\firstpage{1}

\journalname{Nuclear Physics B Proceedings Supplement}

\runauth{A.~Maas}

\jid{nuphbp}

\jnltitlelogo{Nuclear Physics B Proceedings Supplement}

\usepackage{amssymb}
\usepackage{amsthm}
\biboptions{sort&compress}

\usepackage[figuresright]{rotating}

\begin{document}

\begin{frontmatter}

\dochead{}

\title{Observables in Higgsed Theories}

\author{Axel Maas\fnref{support}}

\address{University of Graz, Institute for Physics, Universit\"atsplatz 5, A-8010 Graz, Austria}

\ead{Axel.Maas@uni-graz.at}

\fntext[support]{This work has been supported by the DFG under grant numbers MA 3935/5-1, MA/3935/8-1 (Heisenberg program), and GK 1523/2.}

\begin{abstract}
In gauge theories, observable quantities have to be gauge-invariant. In general, this requires composite operators, which usually have substantially different properties, e.\ g.\ masses, than the elementary particles. Theories with a Higgs field, in which the Brout-Englert-Higgs effect is active, provide an interesting exception to this rule. Due to an intricate mechanism, the Fr\"ohlich-Morchio-Strocchi mechanism, the masses of the composite operators with the same $J^P$ quantum numbers, but modified internal quantum numbers, have the same masses. This mechanism is supported using lattice gauge theory for the standard-model Higgs sector, i.\ e.\ Yang-Mills-Higgs theory with gauge group SU(2) and custodial symmetry group SU(2). Furthermore, the extension to the 2-Higgs-doublet-model is briefly discussed, and some preliminary results are presented.
\end{abstract}


\end{frontmatter}

\section{Introduction}

The standard model Higgs sector contains essentially an SU(2) Yang-Mills theory coupled to two flavors of a complex Higgs field, forming the custodial doublet. This theory therefore has a gauge group SU(2) and a global custodial SU(2) symmetry. Most importantly, the potential of the Higgs field is arranged such as to induce a Brout-Englert-Higgs (BEH) effect, to provide both the Higgs and the gauge bosons with normal masses.

The standard approach is to treat the theory perturbatively \cite{Bohm:2001yx}, in a suitable gauge \cite{Lee:1974zg}. Then, these elementary particles are interpreted as physical degrees of freedom, and especially as observable final states in experiments. Though this is a very successful approach, it poses the question why this is consistent. On the one hand, it is possible to write down gauges in which, e.\ g., the $W$ and $Z$ gauge bosons, remain massless to all orders of perturbation theory \cite{Maas:2012ct}. On the other hand, both the gauge bosons and the Higgs carry the weak charge which, as a non-Abelian charge, is not observable, and, indeed, gauge-dependent \cite{Haag:1992hx}. These particles can therefore not be observable degrees of freedom \cite{Banks:1979fi,'tHooft:1979bj,Frohlich:1980gj}.

Indeed, a more formal treatment \cite{Frohlich:1980gj,Frohlich:1981yi} shows that only composite operators, effectively bound states, could be the appropriate gauge-invariant degrees of freedom. Though this is a conceptually satisfying answer, this still requires to explain why then a perturbative description using the gauge-dependent degrees of freedom is successful in describing the observable spectrum. Especially, the same arguments about unobservability can be made for QCD, where the quarks and gluons are indeed not the appropriate observed states\footnote{The situation in QED is different, due to its Abelian nature \cite{Haag:1992hx}.}.

This is resolved by the Fr\"ohlich-Morchio-Strocchi (FMS) mechanism  \cite{Frohlich:1980gj,Frohlich:1981yi}, and the key element is the BEH effect: Select a suitable gauge, i.\ e.\ one with non-vanishing Higgs expectation value. Considering a bound-state operator in the $J^P=0^+$ quantum number channel, and performing an expansion in the fluctuations $\eta$ of the Higgs field $\phi$ around it vacuum expectation value $v$ yields\footnote{This is a rather subtle statement here, as a $W$-ball operator with the same $J^P$ quantum number would no contribute to leading oder in this expansion.} \cite{Frohlich:1981yi}
\bea
&&\langle\phi_i^\dagger(x)\phi^i(x)\phi_j^\dagger(y)\phi^j(y)\rangle\nn\\
&\approx& v^4+4v^2(c+\langle\eta^\dagger_i(x) n^i n_j^\dagger\eta^j(y)\rangle)+{\cal O}(\eta^3)\label{correl},
\eea
\no where $n$ is the direction in weak isospin space of the Higgs expectation value, and $c$ some constant. Hence, up to this order, the poles of the bound state and the elementary particle coincide, and thus have the same mass. A similar relation holds for the $1^-$ channel of the $W$. However, in that case a twisting occurs, relating the gauge triplet $W$ to a custodial triplet bound state \cite{Frohlich:1981yi}. In a similar manner cross-sections approach in this expansion their perturbative values.

\section{The standard model case}

Though this is a convincing argument theoretically, it is based on an expansion. A check requires non-perturbative methods. One possibility to do so is lattice gauge theory. The minimum requirement to check is just Yang-Mills-Higgs theory, i.\ e.\ (degenerate) $W$ and $Z$ gauge bosons, and the complex Higgs doublet, as described by the Lagrangian
\bea
{\La}&=&-\frac{1}{4}W_\mn^aW^\mn_a+(D_\mu\phi)^\dagger D^\mu\phi\nn\\
&&-\gamma(\phi\phi^\dagger)^2-\frac{m_0^2}{2}\phi\phi^\dagger\label{action}\\
W_\mn^a&=&\pdm W_\nu^a-\pdn W_\mu^a-g f^{abc} W_\mu^b W_\nu^c\nn\\
D_\mu^{ij}&=&\pd_\mu\delta^{ij}-igW_\mu^a\tau^{ij}_a\nn,
\eea
\no which can be simulated using standard methods \cite{Maas:2013aia}. It has then be demonstrated explicitly that relations of type \pref{correl} work \cite{Maas:2012tj,Maas:2013aia}. This supports, in as far as numerical simulations can do so, the validity of this picture.

However, this is not a triviality. Though the quantum phase diagram of the lattice-regularized\footnote{Issues of triviality are neglected here, assuming that a lattice cutoff provides a quantitative close-enough simulation of new physics.} version of the theory \pref{action} is simply connected \cite{Fradkin:1978dv}, the relation \pref{correl} does not hold everywhere. On the one hand, a dramatic change occurs when the mass of the $0^+$ singlet is below the one of the $1^-$ triplet. In this case, the theory becomes QCD-like \cite{Maas:2013aia,Maas:unpublished3,Evertz:1985fc,Langguth:1985dr}. Especially, the $W$ propagator has no longer a physical positive-definite spectral function, and shows no discernible pole structure \cite{Maas:2013aia}, like the gluon in Yang-Mills theory or QCD \cite{Maas:2011se}. Though the Higgs still does \cite{Maas:2013aia}, an expansion like \pref{correl} is no longer possible, since the vacuum expectation value then vanishes in all gauges. This result is also remarkable as perturbatively \cite{Bohm:2001yx}, especially also in leading-order lattice perturbation theory, it should in general be possible to find a weakly-interacting Higgs-like domain with a Higgs lighter than the $W$.

A similar situation arises when the $0^+$ mass crosses the decay threshold for a decay into two $1^-$ particles \cite{Maas:unpublished3}. Though it is (lattice-)perturbatively possible to push the Higgs mass up to several 100 GeV before encountering serious problems \cite{Bohm:2001yx}, this seems not to be possible non-perturbatively: Only scattering states appear to remain in the spectrum. However, at the same time the expansion still works, within uncertainties, for the $1^-$ and the $W$ \cite{Maas:2013aia}. The situation is, however, different than in the QCD-like domain. While there the vacuum expectation value vanishes anyway, here just no stable particle appears to remain in the $0^+$ channels, just scattering states.

Of course, in both cases this may change in other parts of the phase diagram, and the inherent problems due to systematic errors can mask effects, especially in the computationally expensive high-mass regime. Nonetheless, the situation is different enough from the expected one to warrant investigating its consequences for other theories. Especially, as additional Higgs fields appear throughout beyond-the-standard-model phenomenology in several theories, and often with large scale differences between the different operative BEH-effects, like in grand-unified theories. 

\section{2-Higgs-doublet model}

\subsection{General considerations}

One of the simplest cases belonging to this class of theories is the 2-Higgs-doublet model (2HDM) \cite{Lee:1974jb}. It introduces a second Higgs doublet $\varphi$, which may or may not interact directly with the original one. The most general Lagrangian includes a potential up to order four, which can either have a SU(4) custodial symmetry, or can be explicitly broken down to any of its subgroups. In the following, only the case with a breaking to SU(2)$\times$SU(2) at the ultraviolet cutoff will be considered, i.\ e.\ the only admitted terms are two distinct mass terms, two distinct terms of type $\phi^4$ and $\varphi^4$, and a term $\phi^{a\dagger}_i\phi^a_i\varphi^{b\dagger}_j\varphi^b_j$ with gauge indices $a$ and $b$ and custodial indices $i$ and $j$, which couples both doublets, but keeps separate custodial symmetries for both doublets. The explicit breaking SU(4)$\to$SU(2)$\times$SU(2) is therefore entirely mediated by the different tree-level masses and coupling constants of the non-mixing $\phi^4$ and $\varphi^4$ terms. As a consequence, physical states can be distinguished by two independent custodial quantum numbers, besides the usual $J^P$ quantum numbers.

The perturbative treatment of theses theories is rather straight-forward, at least at tree-level. Whether one or both fields develop vacuum expectation values can be adjusted by choosing the five independent coupling constants at will.

The situation becomes far less trivial when considering analogues to the relation \pref{correl}. Considering the case of the $0^+$ full singlet channel, which is the relevant one for the Higgs, there are now two naive correlation functions for bound states appearing, one containing the $\phi$ field, and one containing the $\varphi$ field. If for both of them a Higgs effect is assumed, they both expand to the corresponding propagators\footnote{The mixed one expands to a mixed propagator, which vanishes. However, the bound-state correlator will mix with the other $0^+$ singlet correlators, just as with the $W$-ball mentioned above.}. Hence, these results would imply that two physical excitations should be observable, with the corresponding mases of both Higgs doublets.

Similarly, for the $1^-$ triplet propagator two correlators exist, living in two different custodial triplets, which both expand to the elementary gauge-triplet $W$ propagator, in analogy to the single-doublet case \cite{Frohlich:1981yi},
\bea
&&\langle(\tau^a\phi^\dagger D_\mu\phi)(x)(\tau^a\phi^\dagger D_\mu\phi)(y)\rangle\nn\\
&\approx& \tilde{c}\tr(\tau^a\tilde{n}\sigma^i\tilde{n}\tau^a\tilde{n}\sigma^j\tilde{n})\langle W^i_\mu W^j_\mu\rangle+{\cal O}(\eta_\phi W)\label{correl2},
\eea
and
\bea
&&\langle(\omega^r\varphi^\dagger D_\mu\varphi)(x)(\omega^r\varphi^\dagger D_\mu\varphi)(y)\rangle\nn\\
&\approx& \tilde{d}\tr(\omega^r\tilde{m}\sigma^i\tilde{m}\omega^r\tilde{m}\sigma^j\tilde{m})\langle W^i_\mu W^j_\mu\rangle+{\cal O}(\eta_\varphi W)\label{correl3},
\eea
\no where $\tau$, $\omega$, and $\sigma$ are Pauli matrices with indices $a$, $i/j$, and $r$ in the first custodial group, the gauge group, and the second custodial group, respectively. The $\tilde{c}$ and $\tilde{d}$ are some constants, and the $\tilde{n}$ and $\tilde{m}$ are the SU(2)-valued directions of the vacuum expectation values in weak isospin space. This now poses a problem, as this implies that in both custodial doublets states with a mass of the $W$s would appear, and hence six, rather than three, physical states should be observed in experiments. This would not be the case, if only one of the Higgs fields forms a vacuum expectation value, rather than both. If the second field would hence rather act like scalar matter, this problem does not arise, as the expansion would be void.

\subsection{Lattice simulations}

As such considerations would have quite some impact on the properties, and possible realistic parameters, of 2HDM, it is worthwhile to check this expansion. Considering again only the case of the $W$s, the Higgs, and the additional doublet, it is in addition necessary to restrict to a not too high mass scale for the second doublet, given the available computational resources. However, since the aim is anyhow only a proof-of-principle check, this is viable. The 2HDM model itself has already been investigated in lattice calculations, see e.\ g.\ \cite{Lewis:2010ps}, and found to have a much richer phase structure than the theory with just a single doublet. There, however, attention was focused on the question of a breaking of the custodial symmetries, while here the comparison to the perturbative phenomenology is of central interest.

The simplest idea is to study the situation where only one of the doublets experiences a BEH effect, while the other one does not. That will be attempted to be achieved by exposing one doublet to a BEH potential, while the second one is kept massive, and without self-interaction at the ultraviolet cutoff. The first guess to obtain such a situation is giving one doublet the same bare parameters which produce a reasonably strong BEH effect in the single-doublet case, while giving the second doublet no self-interaction at all, and a large tree-level mass. Given that there are large additive mass renormalizations, it is not necessarily clear, whether this approach will succeed.

Extending the lattice simulations of \cite{Maas:2013aia} for this case by adding a second doublet, and treating it like the first in terms of using a Metropolis update intertwined with the other field updates, is straight-forward, and will be detailed elsewhere \cite{Maas:unpublished}. Of course, in addition a second set of tree-level parameters for the Higgs will be required, which in lattice notation \cite{Maas:2013aia} are the gauge coupling $\beta$, the two hopping parameters $\kappa_\phi$ and $\kappa_\varphi$, the self-interactions $\lambda_\phi$ and $\lambda_\varphi$, and the coupling $\gamma_{\phi\varphi}$, where $\phi$ denotes the usual Higgs field, and $\varphi$ the second doublet. The case of large, and positive, mass-squared at tree-level for the second field is attained by setting $\kappa_\varphi=0.01$, a value (much) smaller than $1/8$, which would be a zero tree-level mass. Setting also its tree-level coupling $\lambda_\varphi=0$, but permitting some coupling between both doublets, $\gamma_{\phi\varphi}=1/2$, and $\beta=2.3$ should naively realize the scenario when setting $\kappa_\phi=0.32$ and $\lambda_\phi=1$.

\begin{figure*}
\centering
\includegraphics[width=0.5\linewidth]{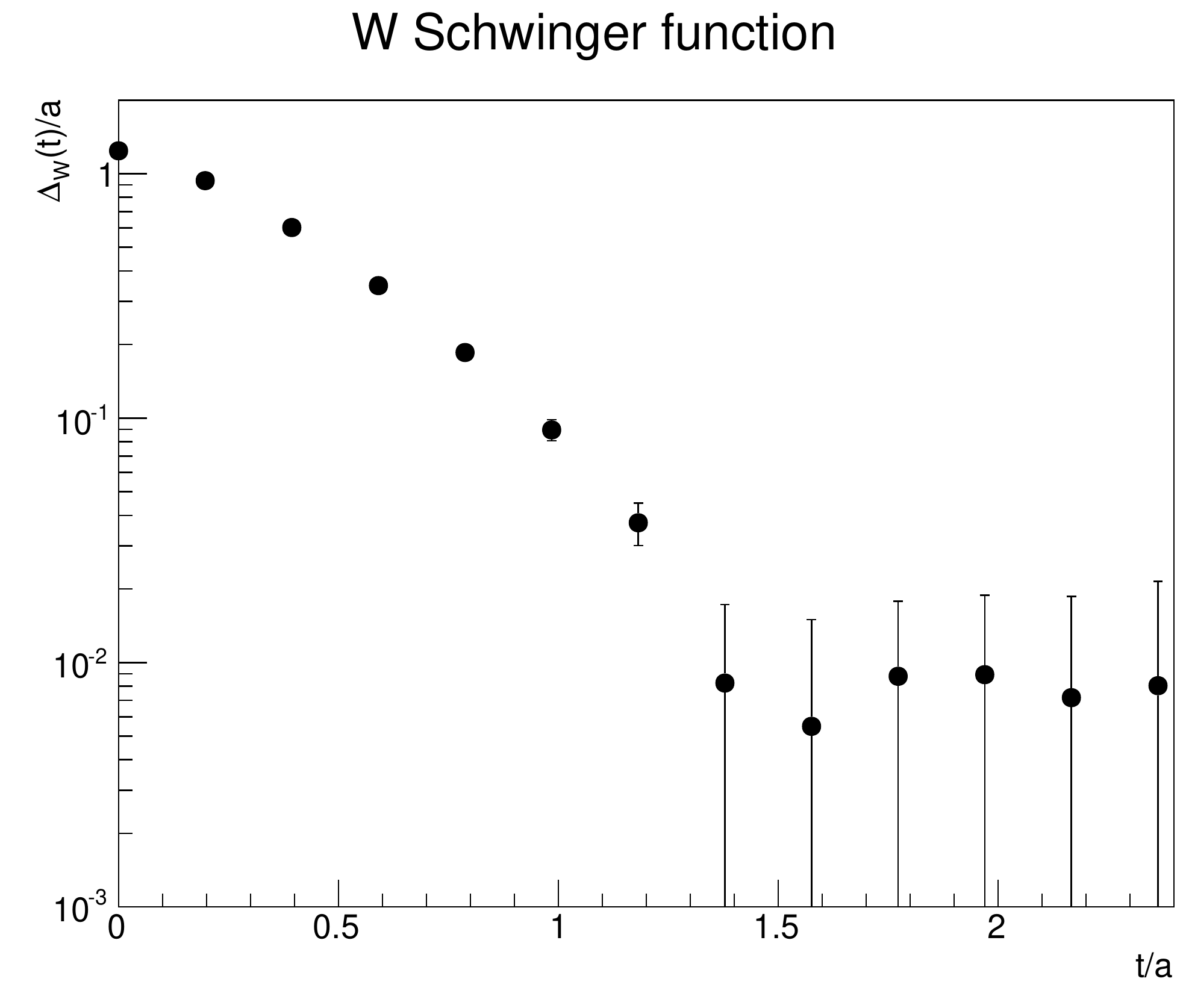}\includegraphics[width=0.5\linewidth]{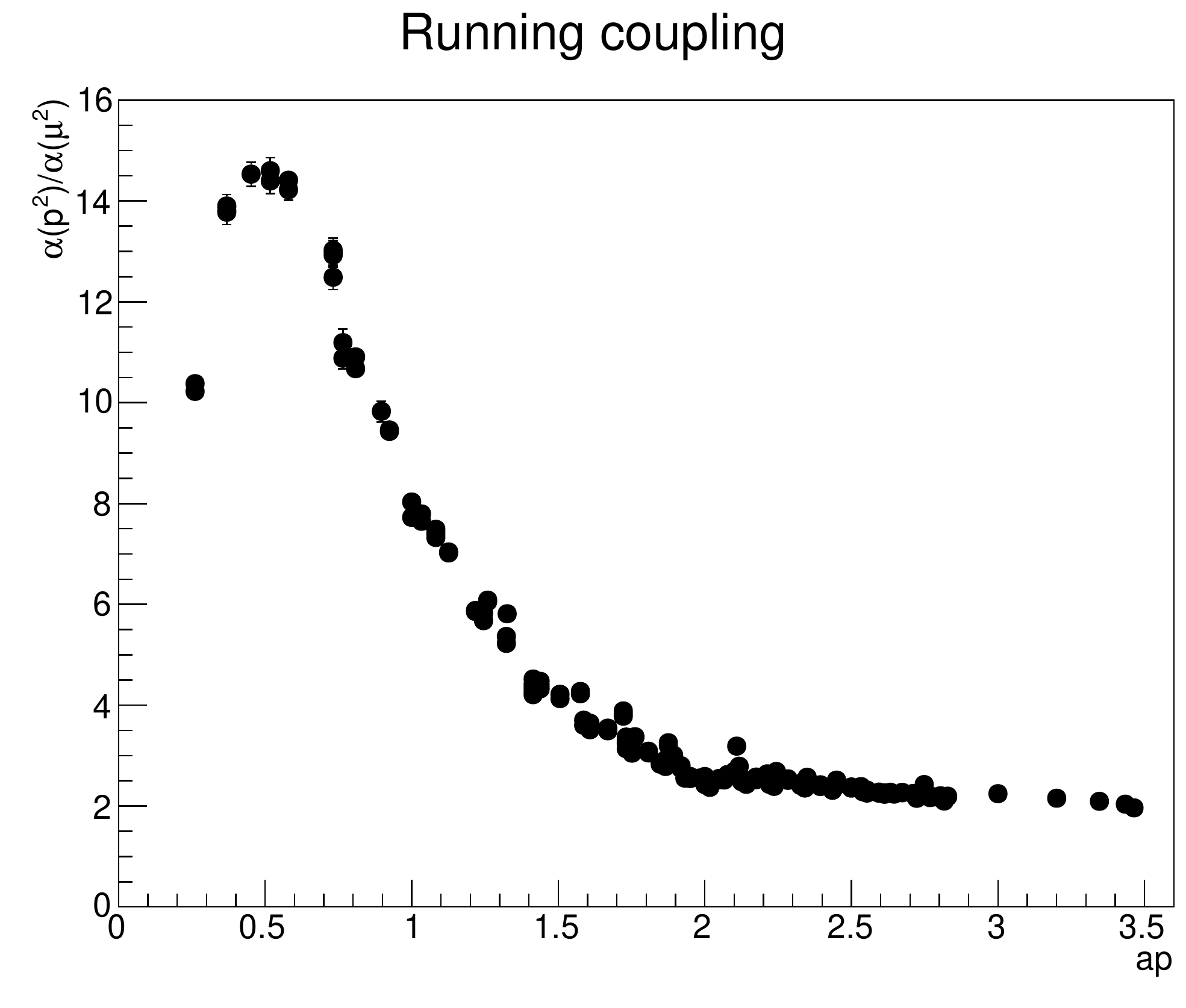}\\
\includegraphics[width=0.5\linewidth]{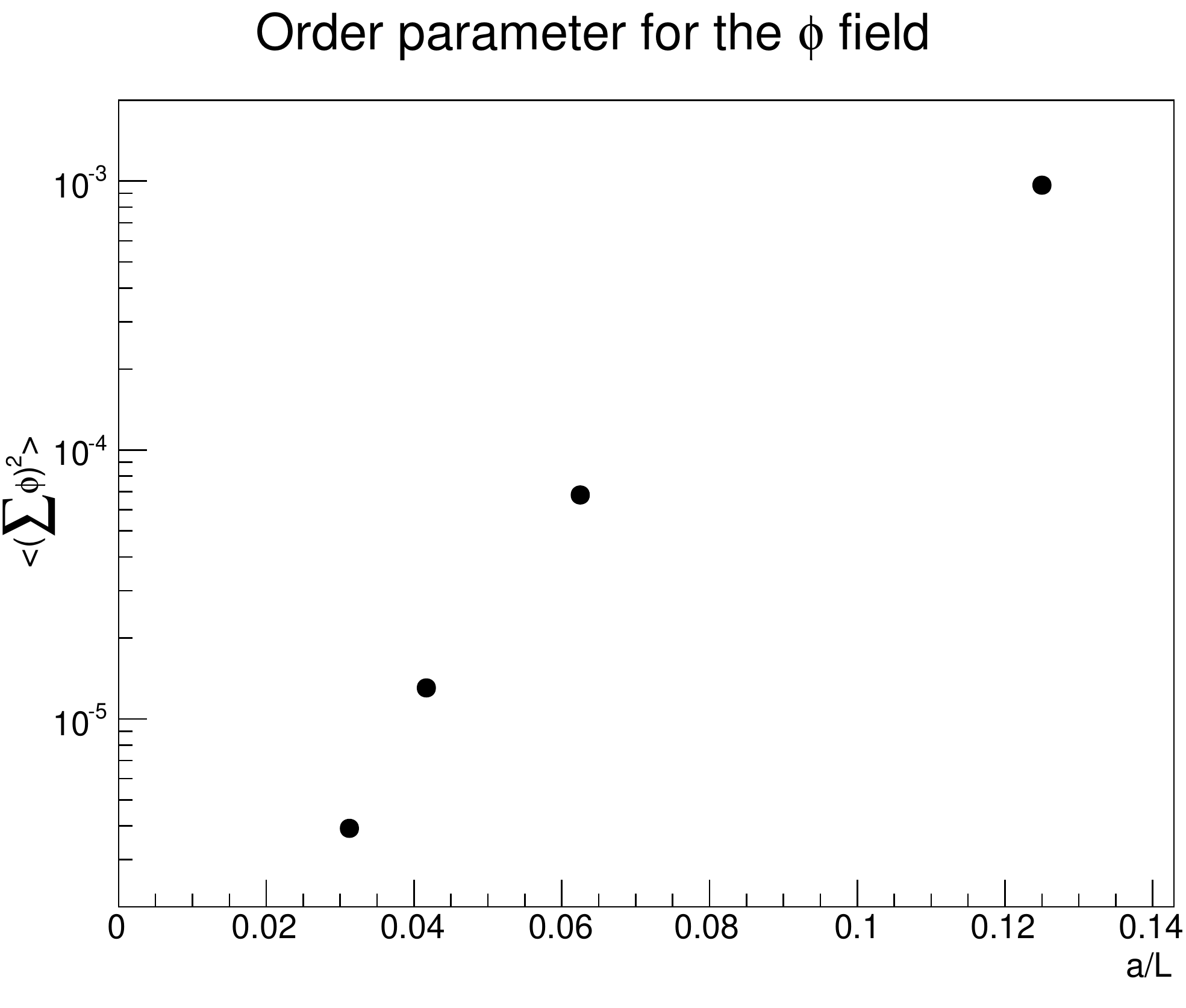}\includegraphics[width=0.5\linewidth]{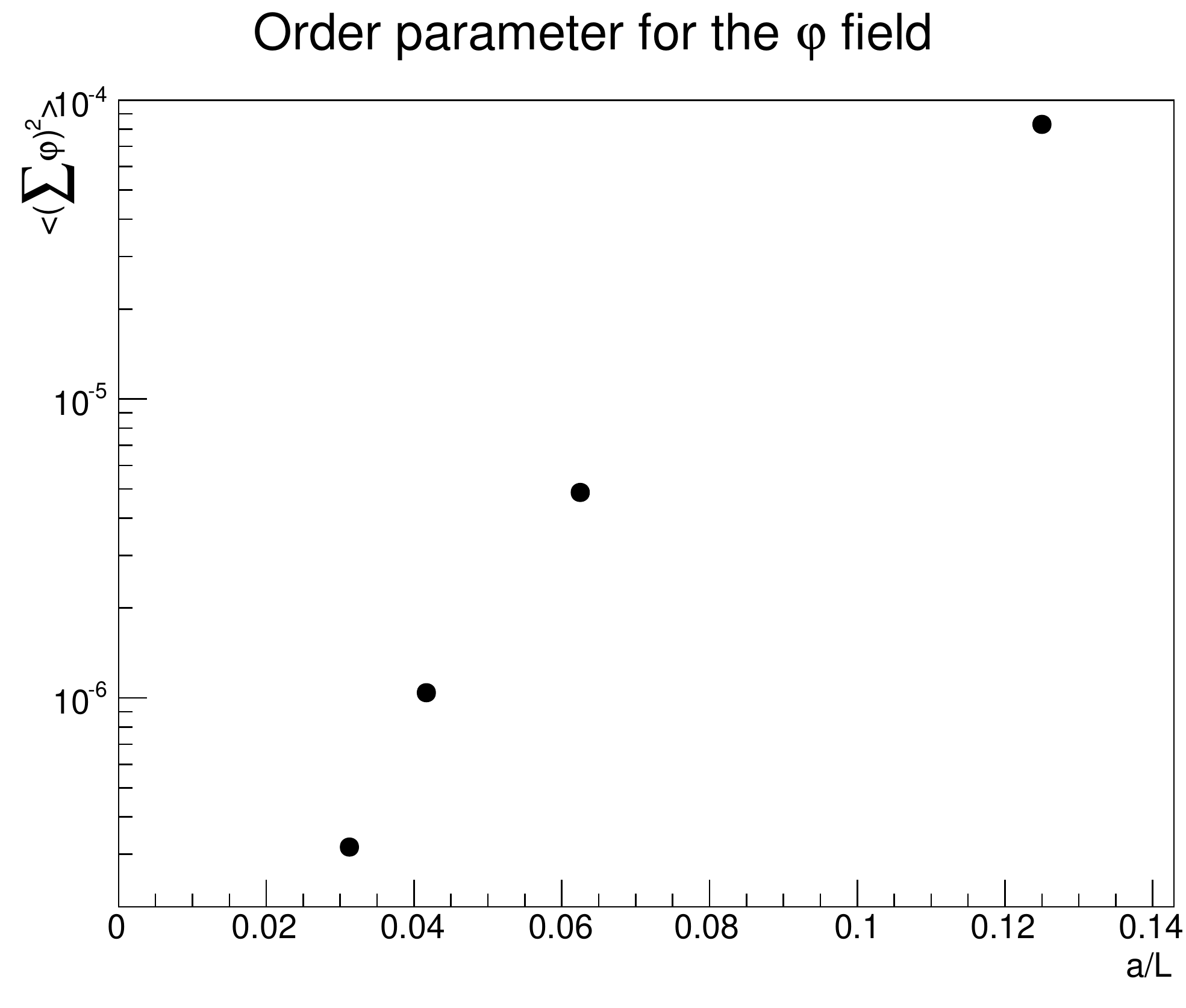}\\
\caption{\label{fig:a}The $W$ propagator in position space (top-left panel), the running gauge coupling in the miniMOM scheme (top-right panel), and the order parameters for the $\phi$ field (bottom-left panel) and the $\varphi$ field (bottom-right panel). Results in the top panels are from a 24$^4$ lattice with the parameters set to the values given in the text. The situation corresponds to a QCD-like behavior.}
\end{figure*}

The results\footnote{The determination of these quantities requires to fix a gauge. For this purpose a non-aligned minimal Landau gauge has been chosen, for technical reasons \cite{Maas:2012ct}. More details on the gauge-fixing procedure and the properties of the gauge can be found in \cite{Maas:2013aia}, as well as a discussion of alternative gauges, and their corresponding advantages and disadvantages. There also details for determining the $W$ propagator and the running gauge coupling can be found.} are shown in figure \ref{fig:a}. That this is not the desired type of physics is visible from the top-left panel already, which shows the the $W$ propagator in position space. It exhibits a clear wrong-sign bending, and has indications for a zero-crossing, signatures of a QCD-like behavior \cite{Maas:2011se}. Also the running gauge coupling in the miniMOM scheme \cite{vonSmekal:2009ae} has a large integrated strength, in contrast to the weak interaction case observed for the same parameters in the case without the second doublet \cite{Maas:2013aia}. In this situation, the theory is not weakly interacting. Finally, the order parameters indicating the metastability characteristic of the BEH effects, defined as the 'relative alignment of the vacuum' \cite{Caudy:2007sf}
\be
v_\phi=\left\langle\left(\sum_x\phi(x)\right)^2\right\rangle\nn,
\ee
\no vanishes with volume in the characteristic way of a theory without BEH effect \cite{Caudy:2007sf} for both doublets. Hence, in this case the theory is indeed very QCD-like. However, the second doublet has not just become light by an additive mass renormalization. When renormalizing the scalar propagator of the second doublet, it turns out that the additive mass shifts are large, while the wave-function renormalization constants are still close to one. This behavior is also observed in the quenched case \cite{Maas:unpublished}, when the tree-level Higgs mass is large, but not if the tree-level Higgs mass is small. This indicates that the intention of a second heavy particles is not spoiled, but the influence of it remains large.

Neither varying just the mass-parameter of the second doublet, nor the properties of the BEH-effect of the Higgs in some range changes this behavior substantially.

A simpler starting point therefore appears to be to have both doublets under the influence of a BEH effect, e.\ g.\ by choosing for the second doublet the tree-level parameters $\kappa_\varphi=0.33$ and $\lambda_\varphi=1.1$, and thus similar to the ones of the Higgs doublet. Also, the tree-level coupling between both doublets is switched off, $\gamma_{\phi\varphi}=0$.

\begin{figure*}
\centering
\includegraphics[width=0.5\linewidth]{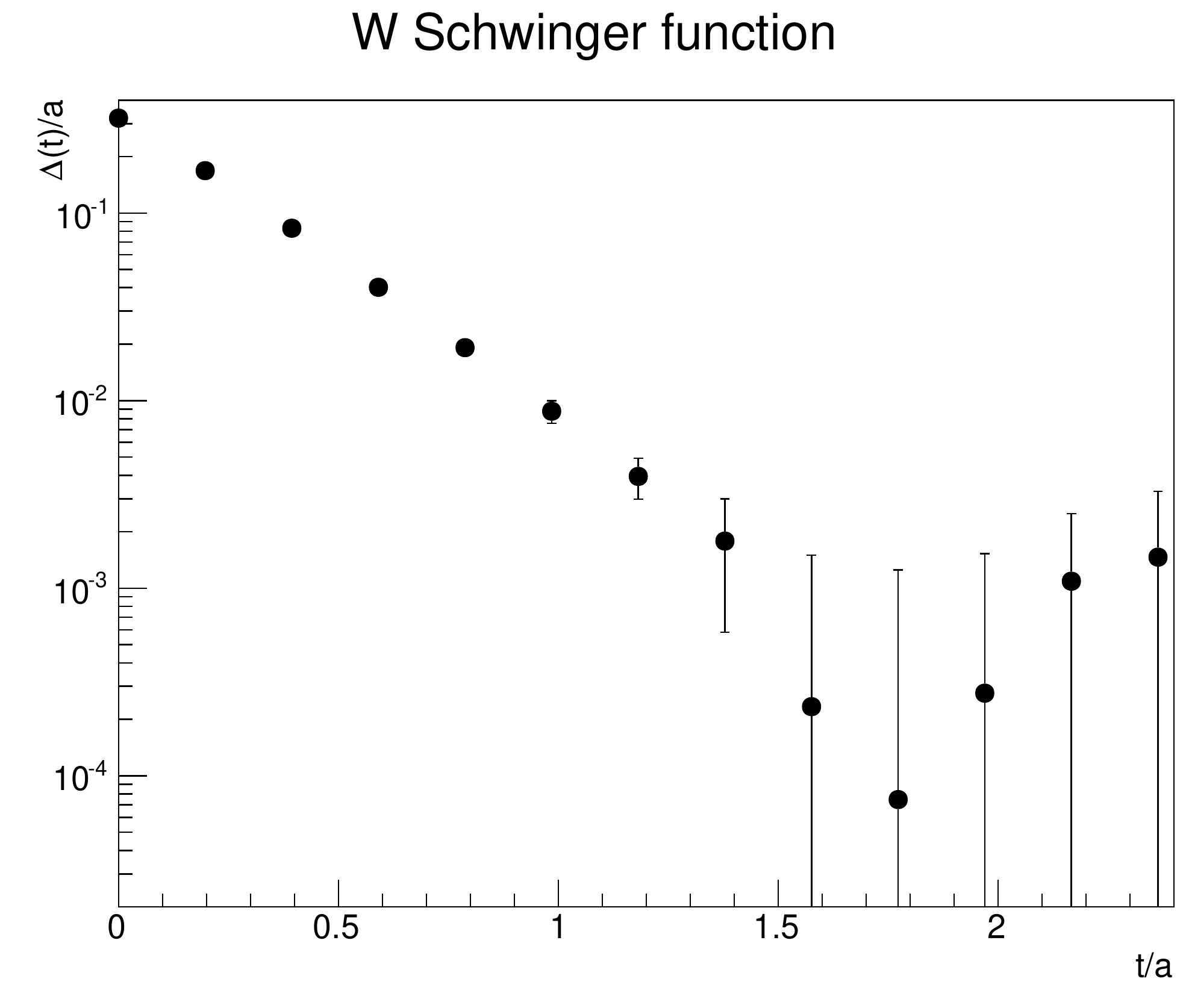}\includegraphics[width=0.5\linewidth]{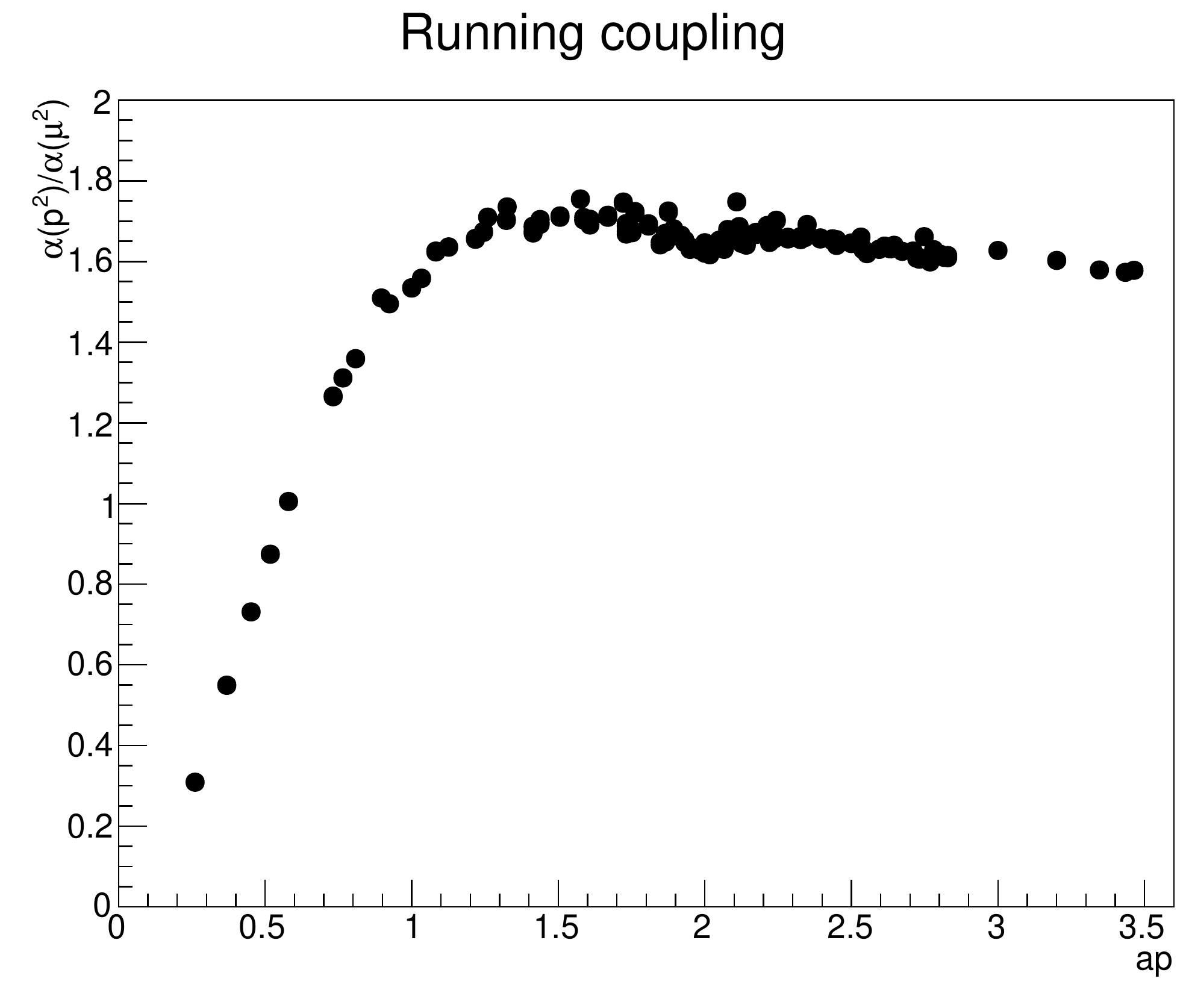}\\
\includegraphics[width=0.5\linewidth]{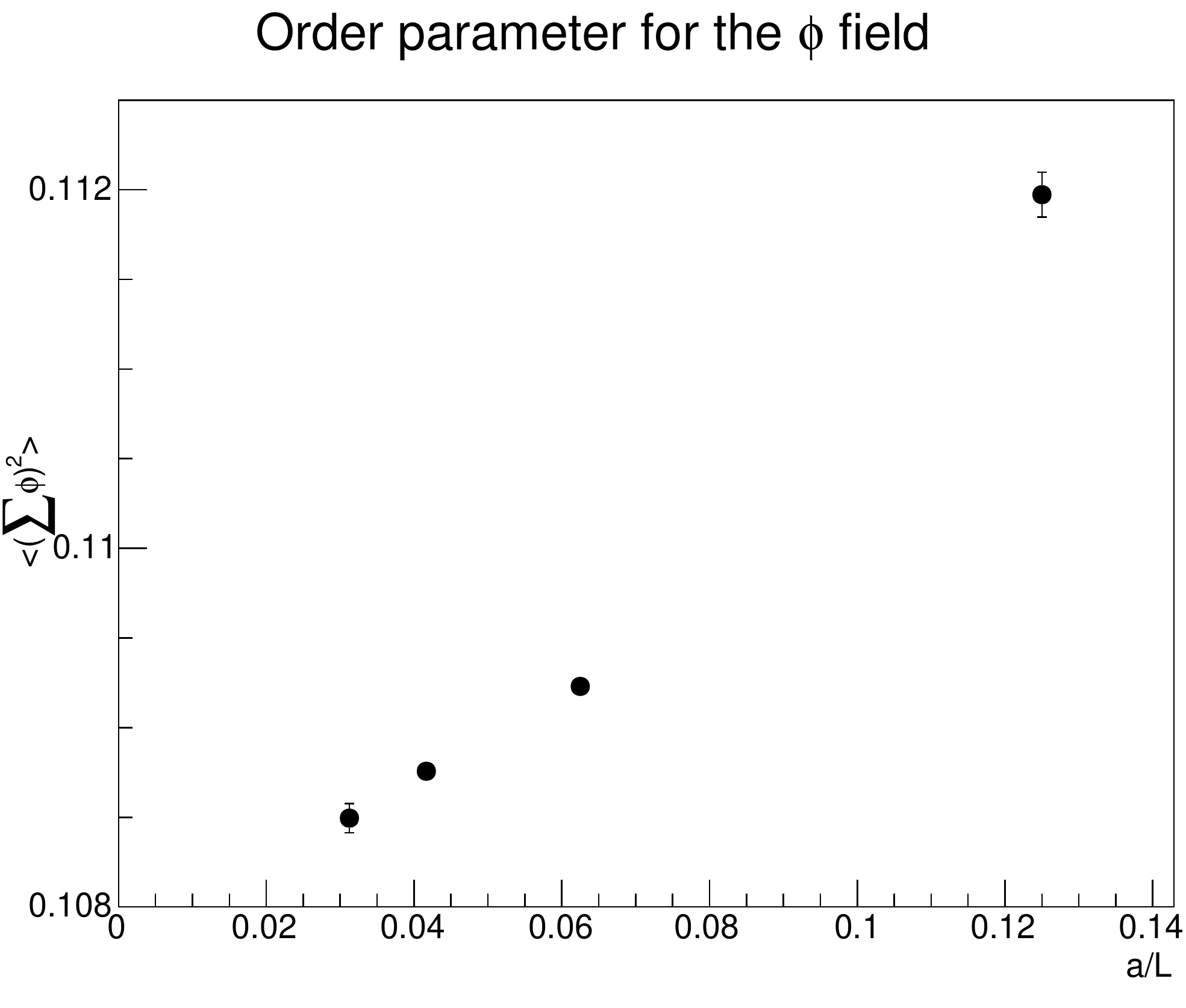}\includegraphics[width=0.5\linewidth]{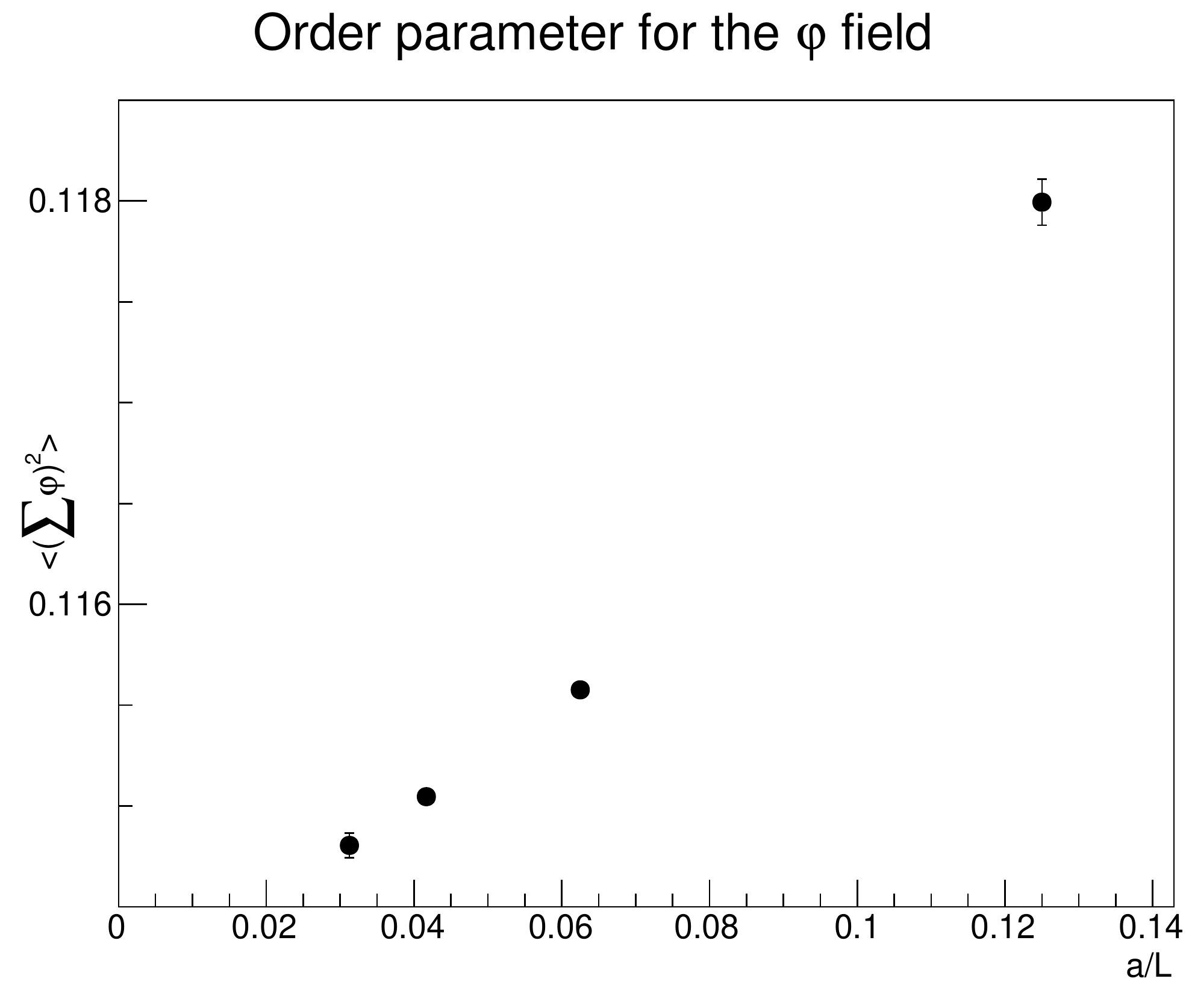}\\
\caption{\label{fig:b}The $W$ propagator in position space (top-left panel), the running gauge coupling in the miniMOM scheme (top-right panel), and the order parameters for the $h$ field (bottom-left panel) and the $H$ field (bottom-right panel). Results in the top panels are from a 24$^4$ lattice with the parameters set to the values given in the text. The situation corresponds to a Higgs-like behavior.}
\end{figure*}

The results are shown in figure \ref{fig:b}. They are drastically different from the case shown in figure \ref{fig:a}. The $W$ propagator shows an, essentially, exponential decay, as in the case with a single doublet when the BEH effect operates \cite{Maas:2013aia}. Also, the integrated strength of the running coupling is now small, corresponding to a weakly interacting theory. Finally, the order parameters are almost constant (note the scales in figure \ref{fig:b}). Since the effective mass of the $W$, as well as the location of the maximum in the running gauge coupling are in general rather good indicators for the scale of the lattice spacing $a$ \cite{Maas:2011se,Maas:2013aia}, this is not an effect of just a poor lattice size or discretization. Hence, in this case both Higgs fields experience a BEH effect.

Interestingly, oddly, enough, this situation seems to be unstable against even a small perturbation in form of a non-zero $\gamma_{\phi\varphi}$, forcing a fall-back into a QCD-like behavior. However, this has to be explored further.

The next steps are now to determine the masses of the two $1^{-}$ custodial triplets, and to check, whether in accordance with the relations \pref{correl2} and \pref{correl3} they are degenerate, or not. If they are, this is of substantial importance for BSM phenomenology. If not, this would indicate that in the present situation the FMS mechanism is not working as in the case of a single Higgs doublet. Finally, the question remains, whether it is possible to find a point where only one doublet experiences a BEH effect. This will require a substantial exploration of the phase diagram.

\section{Summary}

The field-theoretical understanding, and identification, of physical degrees of freedom in the Higgs sector is essentially provided by the FMS mechanism \cite{Frohlich:1980gj,Frohlich:1981yi}, as far as lattice calculations can determine \cite{Maas:2012tj,Maas:2013aia}. This also explains why perturbation theory can be successful, as this is even for an arbitrarily weak interaction by far not an obvious statement \cite{Haag:1992hx}.

Two ingredients enter critically into this result. One is that the $0^+$ custodial singlet state has a mass between one time and two times the $1^-$ custodial triplet, and that the gauge group and the custodial group coincide. If the former is not the case, the FMS mechanism appears to break down, and thus perturbation theory is no longer a good description. If the latter is not fulfilled, the degeneracies of the elementary and physical states is not the same, and thus multiplicities are different from perturbation theory. Fortunately, both conditions are fulfilled in the standard model. Hence, it remains likely that also the application of the FMS mechanism to the full standard model \cite{Frohlich:1980gj,Frohlich:1981yi} appears to be likely correct. However, this must be checked. But especially the presence of parity violation, very different scales, and fermions will require for this most likely off-lattice methods.

A research direction as important is to investigate standard model extension which do, e.\ g.\ due to presence of more Higgs doublets, are not yet covered in the FMS construction. As outlined in the text, already in the case of the 2HDM additional constraints show up. The exploratory investigation of this theory presented here in fact already indicate that naive ideas about its phenomenology could be doubted. This can be extended to other  phenomenologically more relevant theories. Doing this will put both phenomenology and the possibility to generalize the FMS mechanism to the test. Whatever will be the result of this, it will certainly provide new insights into more general theories with BEH effect.

\bibliographystyle{elsarticle-num}
\bibliography{../bib/bib.bib}

\end{document}